\documentclass[10pt,letterpaper]{article}
\usepackage{opex3} 
\usepackage{geometry}
\usepackage{color,graphicx}
\usepackage{amsmath,amssymb,xspace}
\usepackage{array}
\usepackage{float}
\usepackage{cite}
 
\begin{document}


\title{Local ensemble transform Kalman filter, a fast non-stationary control law for adaptive optics on ELTs: theoretical aspects and first simulation results}

\author{Morgan Gray,$^{1,*}$ Cyril Petit,$^{2}$ Sergey Rodionov,$^{1}$ Marc Bocquet,$^{3,4}$\\
Laurent Bertino,$^{5}$ Marc Ferrari$^{1}$ and Thierry Fusco$^{1,2}$}

\address{$^1$Aix Marseille Universit\'e, CNRS, LAM (Laboratoire d'Astrophysique de Marseille)\\ UMR 7326, 13388 Marseille, France\\
$^2$ONERA, 29 avenue de la Division Leclerc, 92322 Ch\^atillon, France\\
$^3$Universit\'e Paris-Est, CEREA joint laboratory \'Ecole des Ponts ParisTech and EDF R\&D,\\
6-8 avenue Blaise Pascal, 77455 Marne la Vall\'ee, France\\
$^4$INRIA, Paris Rocquencourt research center, France\\
$^5$NERSC, Thormohlens gate 47, N-5006 Bergen, Norway}

\email{$^{*}$morgan.gray@lam.fr}


\begin{abstract}
We propose a new algorithm for an adaptive optics system control law, based on the Linear Quadratic Gaussian approach and a Kalman Filter adaptation with localizations. It allows to handle non-stationary behaviors, to obtain performance close to the optimality defined with the residual phase variance minimization criterion, and to reduce the computational burden with an intrinsically parallel implementation on the Extremely Large Telescopes (ELTs).
\end{abstract}

\ocis{(010.1080) Active or adaptive optics; (010.1330) Atmospheric turbulence.}




\section{Introduction}

The prospect of ELTs (GMT \cite{Johns-p-12},TMT \cite{Stepp-p-12}, E-ELT \cite{McPherson-p-12}) with their related high-dimensional Adaptive Optics (AO) systems, such as eXtreme AO (XAO), MultiConjugate AO (MCAO) or even Multi-Object AO (MOAO), has aroused many developments in computationally efficient control algorithms. A good summary of the various techniques recently developed and their references can be found in \cite{Gilles-a-13}. Linear Quadratic Gaussian (LQG) regulator is one of them. This control approach, though introduced a long time ago in AO \cite{Paschall-a-93}, has truly found a resonance only during the last decade. This success is motivated by the optimal control solution provided by LQG, as well as its ability to account naturally for various perturbations such as vibrations \cite{Petit-a-08,Sivo-p-13,Guesalaga-a-12}, limitations such as saturations \cite{Kulcsar-p-07} or Deformable Mirror (DM) dynamics \cite{Correia-a-10}.\\
\indent LQG has thus demonstrated improved performance compared to other control solutions, both in numerical simulations and experimental implementations. However, this control solution suffers from various limitations, the major one being the computational complexity especially in the non-stationary framework. Standard and brute force application of LQG is known to exhibit deterring computational cost, both in terms of off-line computations of the Kalman gain and on-line computations through multiple Matrix Vector Multiplications (MVM). Basically, assuming a number $n$ of state variables in the system, complexity of the off-line computations is proportional to $n^{3}$, and complexity of the on-line computations is proportional to $n^{2}$. Moreover, this number $n$ increases with the square of the telescope diameter in classical AO, and also with the number of reconstructed layers in tomographic approaches. As a consequence, reduction of computational complexity of Kalman Filter (KF)-based control solution has triggered much work in the recent years. The computational issue is significantly increased once considering non-stationary systems. As turbulence and system characteristics evolve with time, one would wish to update the control solution accordingly. For the LQG, this means redoing the off-line computations regularly, hence an additional computational cost.\\
\indent To circumvent these limitations, the Ensemble Transform Kalman Filter (ETKF) has been previously introduced \cite{Gray-p-12}. This approach derives from methods developed in geophysics and meteorology in order to adapt the KF to large scale systems with non-stationary models. The ETKF does indeed provide a theoretical alternative to the KF for non-stationary systems but has also dramatic computational limitations for these large scale systems. Thus, a new control solution is discussed in which ETKF has been modified to implement a zonal and localized approach briefly proposed in \cite{Gray-p-13}. This present paper differs from the previous one in that we are giving a detailed description of this new approach and the required mathematical formalism with new simulations in the case of a 16 m telescope and the resolution of the differential pistons issue. The zonal approach is motivated by the sparse matrices involved as well as by the difficulties encountered when dealing with huge number of modes (typically, a few thousand to tens of thousands for ELT AO systems) such as Zernike or Karhunen-Lo\`{e}ve modes (edge effects, numerical issue in modes computing). In addition, zonal approach also allows to use Fourier domain decomposition and WF reconstruction, benefiting from fast computation, though Fourier domain based approaches may suffer from edge effects on circular (annular) apertures, requiring particular handling \cite{Poyneer-a-02,Correia-p-08}. Localization here stands for decomposition of the system pupil into small domains on which local estimations are performed. All the local estimations are only combined in a final step, in order to deduce the full-pupil information. The result is thus a spatially distributed estimation based on the ETKF, leading to a hierarchical control scheme. This approach allows a highly scalable implementation, as local computations are done with a reduced complexity and can be performed by a dedicated parallelized computation resource (CPUs/GPUs cluster).\\
\indent This article is organized as follows. Section 2 recalls the main ideas behind using the LQG control and a KF for a classical AO system, its limitations and some solutions developed to overcome these drawbacks. Section 3 explains why the ETKF-based control law can handle non-stationary behaviors and can offer the optimality of the KF, but becomes unsuitable for computational reasons. In section 4 we present a new algorithm, the Local ETKF, and we explain why it is highly suitable for classical AO systems on ELTs (fast and naturally parallel implementation, non-stationarity). In order to assess the theoretical analysis and to demonstrate the potentiality of this new control law, some numerical simulations are detailed in section 5. Finally, conclusion and outlook of our work in progress are given in section 6.


\section{The classic LQG control solution: an optimal control law for AO}

It is now well-known that LQG provides the optimal control solution in AO, at least as long as the various underlying models (turbulence, AO system components, noise) are consistent with reality. To some extent, models can be approximated and LQG proves to be more efficient than any other control solution and to be also robust. Usually, it is still pointed out that this solution suffers from higher complexity and computational cost. Various works have been carried out to reduce this complexity and to propose faster computation even for large scale AO systems. As the Local ETKF basically derives from KF and uses similar formalism, in the following we first recall the basics of LQG and point out its main limitations. For the sake of simplicity, we discuss hereafter the LQG implementation in the framework of a classical AO system, although extension to tomographic systems has already been addressed \cite{Gilles-a-13,Costille-a-10}.

\subsection{Optimality criterion, discrete-time equivalence}

Let us consider the simple AO system described in Fig. 1. The phases $\phi^{\rm tur}(t)$, $\phi^{\rm cor}(t)$ and $\phi^{\rm res}(t)$ represent respectively the incoming turbulent phase, the AO correction provided by the DM and the residual phase.
\begin{figure}[htbp]
\centering\includegraphics[width=10cm]{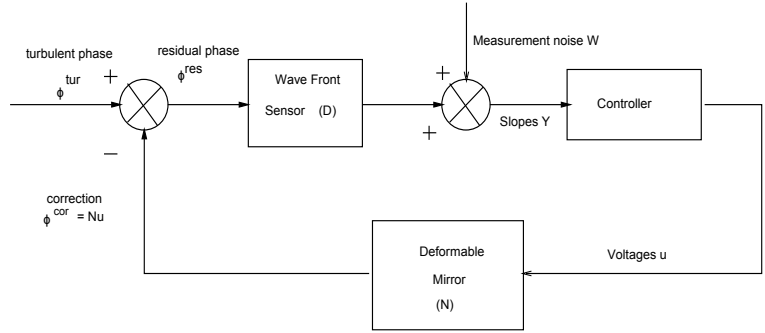}
\caption{Block diagram of a classical AO system closed-loop.}
\end{figure}
In the astronomical framework considered here, the performance criterion consists in minimizing the variance of the residual phase (identified usually as the empirical variance computed over a sufficiently large amount of time), with respect to the DM controls $u$. Though turbulence is a continuous time phenomenon, the residual phase is usually integrated by the WaveFront Sensor (WFS) over a time period $\Delta T$ (frame), the AO control system is digital and correction is applied to the DM  through a Zero Order Hold (ZOH) so that controls are constant over time period $\Delta T$: see chronogram of the AO system in Fig. 2.
\begin{figure}[htbp]
\centering\includegraphics[width=10cm]{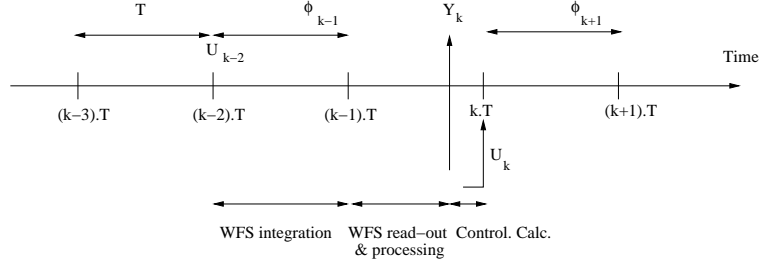}
\caption{Temporal diagram of the system process.}
\end{figure}
In this context, Kulcs\'ar has shown in \cite{Kulcsar-a-06} that the optimal turbulence correction by a sampled AO system is a problem that can be fully addressed in discrete-time. Defining for any continuous-time variable $x(t)$ its discrete-time counterpart $x_{k}$ as:
\begin{equation}
x_{k} {\buildrel\triangle\over =} \frac{1}{\Delta T}\int_{(k-1)\Delta T}^{k\Delta T}x(t)\rm dt ,
\end{equation}
the optimal performance criterion comes down to minimizing the discrete-time criterion:
\begin{equation}\label{Crit_Opt}
J(u) {\buildrel\triangle\over =} \lim_{n\to+\infty}\frac{1}{n}\sum_{k=1}^{n}||\phi^{\rm res}_{k}||^{2}.
\end{equation}
\indent This equivalence is independent from the control solution, and while this result has been demonstrated assuming infinite dynamics of the DM (instantaneous response) in \cite{Kulcsar-a-06}, Correia has also extended it to the case of non negligible DM dynamics in \cite{Correia-a-10}.\\
\indent As a conclusion, in the following we will consider the optimal criterion defined by Eq. \eqref{Crit_Opt}, and we will consider for that a discrete-time control problem. We will thus assume that the WFS is linear and provides a measurement $y_{k}$ of the incoming wavefront $\phi_{k}$, integrated over the frame period $\Delta T$, plus additional noise through the relation:
\begin{equation}\label{Eq_Obs}
y_{k} = {\rm D}\phi_{k-1} + w_{k},
\end{equation}
where D is a linear operator accounting for the WFS measurement of the phase and $w_{k}$ is a discrete zero-mean Gaussian measurement white noise. For the sake of simplicity, we will also assume that the DM is linear, has no dynamics and can be described by its influence matrix N (the abscence of DM dynamics is clearly an optimistic hypothesis, though valid in most current AO systems: the impact of DM dynamics in the proposed control scheme is beyond the scope of this paper and should be addressed in the future). Its correction over a frame period $\Delta T$ is constant (due to the ZOH) and is therefore equal to:
\begin{equation}\label{Eq_Cor}
\phi^{\rm cor}_{k} = {\rm N}u_{k-1}.
\end{equation}
A two-frame delay AO system is thus considered as described in Fig. 2.\\
\indent Now, the AO control problem of minimizing the discrete-time criterion (Eq. \eqref{Crit_Opt}) in order to provide the optimal AO correction of the incoming turbulent wavefront $\phi^{\rm tur}(t)$ finds its solution using the stochastic separation theorem \cite{BarShalom-a-74}. In a first step, the turbulent phase is estimated. The solution to this stochastic minimum-variance estimation problem finds its expression in a KF. Then, the estimated phase is projected onto the mirror's space with a simple least-square solution. The overall control scheme is thus a typical LQG regulator.

\subsection{Mathematical formulation for a classical AO system}

By using the structure of the LQG control and the associated KF described in \cite{Kulcsar-a-06,Petit-a-09}, the turbulent phase ${\varphi}^{\rm tur}$ can be defined on a zonal or a modal basis, and $x_{k}$, the state vector at time $k$, contains 2 occurences of this turbulent phase and 2 occurences of the voltages $u_{k}$:
\begin{equation}\label{Def_xk}
x_{k}=\left(\begin{array}{cccc}{(\varphi_{k}^{\rm tur})^{\rm T}} & {(\varphi_{k-1}^{\rm tur})^{\rm T}} & {(u_{k-1})^{\rm T}} & {(u_{k-2})^{\rm T}}\end{array}\right)^{\rm T}.
\end{equation}
The \textit{stationary} stochastic linear state-space model can be defined with these two equations:
\begin{equation}\label{Syst_mod}
\begin{split}
&x_{k+1} = {\rm A} \times x_{k} + {\rm B} \times u_{k} + v_{k}\\
&y_{k} = {\rm C} \times x_{k} + w_{k}.
\end{split}
\end{equation}
The first equation characterizes the dynamics of the turbulent phase described for instance by a first-order Auto-Regressive (AR 1) model (a higher order AR model could
be also considered):
\begin{equation}
x_{k+1} = \left(\begin{matrix}
{\rm A_{\rm tur}} & 0 & 0 & 0 \\ {\rm Id} & 0 & 0 & 0 \\ 0 & 0 & 0 & 0  \\ 0 & 0 & {\rm Id} & 0 \\
\end{matrix} \right) x_{k}+\left(\begin{matrix}
   0  \\ 0  \\ {\rm Id} \\ 0  \\
\end{matrix} \right) u_{k} + \left(\begin{matrix}
   {\rm Id}  \\ 0  \\ 0  \\ 0  \\
\end{matrix} \right) v_{k}.
\end{equation}
The vector $u_{k}$ contains the voltages that are applied on the DM. The vector $v_{k}$ is a zero-mean white Gaussian model noise. The second equation is the observation equation:
\begin{equation}
y_{k} = \left[\begin{matrix}
0 & {\rm D} & 0 & - {\rm D\times N}\\ \end{matrix} \right] x_{k} + w_{k}.
\end{equation}
The vector $y_{k}$ contains the measurements of the residual phase $\varphi_{k}^{\rm res} = \varphi_{k}^{\rm tur} - \varphi_{k}^{\rm cor}$. As we wrote in section 2.1, the estimation part can be separated from the control part. The optimal control $u_{k}$ can be obtained by separately solving a deterministic control problem and a stochastic minimum variance estimation problem. In Eq. \eqref{Def_xk}, the state vector $x_{k}$ includes control voltages, but these components do not need estimation. Thus, in the following, the vector $x_{k}$ contains \textit{only} the 2 occurences of the tubulent phase. The symbol \textit{k / k} denotes \textit{analysis} or \textit{update} and the symbol \textit{k / k-1} denotes \textit{forecast} or \textit{prediction}. In the linear Gaussian case, the optimal solution of the update estimate of the turbulent phase ${\varphi}^{\rm tur}$ is given by a KF:
\begin{equation}\label{Est_maj}
\hat{x}_{k/k} = \hat{x}_{k/k-1} + H_{k}(y_{k} - \hat{y}_{k/k-1}),
\end{equation}
where $\hat{y}_{k/k-1} = C\times\hat{x}_{k/k-1}$, and with the update estimation error covariance matrix:
\begin{equation}\label{Cov_maj}
\Sigma_{k/k} = ({\rm Id} - H_{k}{\rm C_{1}})\Sigma_{k/k-1},
\end{equation}
and the Kalman gain:
\begin{equation}\label{K_Gain}
H_{k} = \Sigma_{k/k-1}{\rm C_{1}^{T}}({\rm C_{1}}\Sigma_{k/k-1}{\rm C_{1}^{T}} + \Sigma_{w})^{-1}.
\end{equation}
$\Sigma_{w}$ is the measurement noise covariance matrix and $C_{1}$ is an extracted matrix from C, defined by $C_{1} = \left[\begin{matrix} 0 & {\rm D} \end{matrix} \right]$. The prediction estimate, at time k, is simply obtained with $\hat{x}_{k+1/k} = A_{1}\hat{x}_{k/k}$ where $A_{1}$ is an extracted matrix from A, defined by $A_{1} = \left(\begin{matrix} {\rm A_{\rm tur}} & 0 \\ {\rm Id} & 0 \\ \end{matrix}\right)$. The prediction estimation error covariance matrix is calulated through the Discrete Algebraic Riccati matrix Equation (DARE):
\begin{equation}\label{Cov_pre}
\Sigma_{k+1/k} = {\rm A_{1}}\Sigma_{k/k-1}{\rm A_{1}^{T}} - {\rm A_{1}}H_{k}{\rm C_{1}}\Sigma_{k/k-1}{\rm A_{1}^{T}} + \Sigma_{v},
\end{equation}
where $\Sigma_{v}$ is the model noise covariance matrix. The deterministic control problem is simply solved through a least-squares projection of the predicted phase $\hat{\varphi}_{k+1/k}^{\rm tur}$ (upper part of $\hat{x}_{k+1/k}$) onto the DM's space. With a basic KF implementation, we must solve the DARE in order to calculate this Kalman gain $H_{k}$ \cite{Kulcsar-a-06,Petit-a-09}. As we defined a stationary model, all the matrices of the state-space model are time-constant during the observation and the asymptotic formulation of the KF is applied without loss of optimality: the Kalman gain is then precalculated with an off-line computation by solving the DARE, which gives the asymptotic Kalman gain ($H_{k} = H_{\infty}$).

\subsection{LQG limitations for AO systems on ELTs}

LQG has proved to provide efficient control and improved performance compared to other control solutions in the numerical simulations as well as in the experimental validations. Since the very first works in AO \cite{Paschall-a-93}, various developments have been done in classical AO \cite{Kulcsar-a-06,LeRoux-a-04,Looze-a-09} and in wide field tomographic AO \cite{Petit-a-09,LeRoux-a-04,Gilles-a-05,Piatrou-a-07}. Experimental validations in laboratory have been also carried out both in AO \cite{Petit-a-08,Petit-a-09} and MCAO \cite{Costille-a-10,Parisot-t-12}, while first validations on sky \cite{Sivo-p-13,Petit-p-08,Guesalaga-a-13} provided good demonstration of performance.\\
\indent Nevetherless this control solution suffers from various limitations, the major one being the computational complexity, especially in the non-stationary framework.

\subsubsection{Numerical cost and computational complexity with stationary models}

As recalled in introduction, brute force implementation of LQG is prohibitive in terms of computational complexity. As a consequence, reduction of computational cost of KF-based control solution has drawn much work in the recent years. For instance, Poyneer \cite{Poyneer-a-07} has proposed the use of Fourier decomposition of atmospheric turbulence and the statistical independence of the Fourier modes to define a mode-by-mode regulator. A modal KF is used. This approach benefits naturally from mode decoupling (sparsity) and possible parallelization leading to computation speed up. Correia \cite{Correia-p-10} improved the off-line computation of the Kalman gain and real time operations by replacing MVM by spectral iterative methods with sparse approximation of the turbulence covariance matrix which avoids the resolution of the DARE. Massioni \cite{Massioni-a-11} proposed the Distributed Kalman Filter (DKF), which approximates the Kalman gain by assuming the telescope pupil as the cropped version of an infinite-sized phase screen. The method is based on a Fourier transform that is used to decompose an infinite-order system into an infinite set of low, finite-order ones. Fourier transform is used as an off-line tool to accelerate Kalman gain computation, but on-line computation is also improved by sparsity in a zonal representation of phase. Gilles compares in \cite{Gilles-a-13} the performance and costs of two computationally efficient Fourier based tomographic wavefront reconstruction algorithms, the DKF and the iterative Fourier Domain Preconditioned Conjugate Gradient combined with a Pseudo-Open-Loop Control, showing very good performance of DKF and an acceptable computational burden. At the end, most of these solutions tends to propose a drastic reduction of the numerical complexity which becomes $O(n\times log(n))$.

\subsubsection{Dealing with non-stationary turbulence}

LQG (but also control approaches that do rely on turbulence statistics models) is usually or systematically derived in a time-invariant framework: the turbulence model and the system models (WFS, DM responses) are considered as time-invariant. The turbulent model is usually pre-defined and the model matrices are usually derived from information gathered on the spot right before observation \cite{Sivo-p-13}, or deduced from global statistics. This leads obviously to significant simplifications, both in formalism and computation: in particular, one can compute an asymptotic Kalman gain, as mentioned before, which significantly reduces the real-time complexity of LQG. This is also motivated by a pragmatic trade-off between accuracy and efficiency: the goal is to provide good performance with simple enough models of turbulence, thus usually mentioned as control-oriented models. Solutions mentioned in the previous section in order to speed up off-line and on-line computations rely on a time-invariant framework.\\
\indent While we can expect system models to present slow time evolution (or at least predictible), this can not be said about turbulence models as turbulence statistics (seeing conditions, average wind profile ...) clearly evolve over time. Then, one can consider updating the turbulence models, once in a while, based either on dedicated instruments information (seeing monitor) or real-time data as it is already performed on the SPHERE AO system for Tip-tilt control \cite{Petit-p-08, Meimon-a-10}. In the latter case, closed-loop data are directly used to identify periodically the turbulent Tip-tilt models as well as possible vibrations. The resulting identification is used to update the Kalman models and control solution. However, this implies a full update of turbulence models and recomputation of the Kalman gain, as well as correct management of transitions \cite{Guesalaga-a-13}. While this is affordable on limited-dimensional systems such as SPHERE, considering large scale AO systems comes to a dead-end. Recent developments of fast computation of Kalman gain \cite{Correia-p-10,Massioni-a-11} may provide an interesting way out.\\
\indent However time-invariant system is not a prerequisite, and time-variant LQG can also be derived similarly: in the second case, the turbulent model matrices are time-variant, and consequently the Kalman gain computation can not be precomputed off-line by convergence of the DARE (Eq. \eqref{Cov_pre}). Therefore, it must be computed at each step based on the current values of the turbulent model and system matrices through Eqs. \eqref{K_Gain} and \eqref{Cov_pre}. Of course this situation leads to an unsuperable computational complexity as well as to the problem of defining, on every other steps, these time-evolving matrices. Thus, embedding turbulence models identification and update within the control scheme, without significant increase of computational cost while ensuring performance, still represents an unreachable Graal.\\
\indent The Local ETKF-based solution aims both at proposing a time-variant KF-based control solution and a reduced complexity, benefiting from the intrinsically parallel algorithm.


\section{The ETKF: large scale systems adaptation and non-stationary models possibility}

Although the LQG approach for AO systems with a stationary model has been successfully implemented on 4-8 m class telescopes, a similar implementation with non-stationary models is not possible for complex AO systems on ELTs. The major reason is the transition to very high-dimensional systems when explicit storage and manipulation with theoretical covariance matrices are not possible. In the KF-based method, there are two difficulties in inverting $({\rm C_{1}}\Sigma_{k/k-1}{\rm C_{1}^{T}} + \Sigma_{w})$ in Eq. \eqref{K_Gain}. The first one is the size: it is a $p\times p$ matrix (p is the number of measurements). For complex AO systems on ELTs, it can be costly to compute the inverse when $p = O(10^{5})$. The second one is the fact that this matrix will be ill-conditioned and it may be extremely difficult to accurately evaluate its inverse. This situation led us to adapt a new method for large scale AO systems with non-stationary models.

\subsection{Main ideas and mathematical formulation}

The original Ensemble Kalman Filter (EnKF) method is based on the KF's Eqs. \eqref{Est_maj} and \eqref{Cov_maj}, except that the Kalman gain (Eq. \eqref{K_Gain}) is calculated from the covariance matrices provided by an ensemble of \textit{model states}: it is a Monte Carlo method \cite{Burgers-a-98,Evensen-a-03}. A fixed number $m$ of model states of the turbulent phase on the whole pupil of the telescope composes the members of the ensemble $\mathcal{X} = \{x^{1};...;x^{m}\}$, and by integrating it foward in time, it is possible to calculate the empirical estimation error covariance matrices with the following statistical estimator:
\begin{equation}\label{Cov_emp}
\Sigma = Z \times Z^{\rm T} \qquad {\rm with} \qquad Z = [x^{1}-\overline{x};...;x^{m}-\overline{x}]/\sqrt{m-1},
\end{equation}
where $\overline{x} = \frac{1}{m}\sum\limits_{i = 1}^{m}{x}^{i}$ is the mean of the ensemble's members and $Z$ is called the matrix of the \textit{anomalies}. To be precise, $Z_{k/k-1}$ is the prediction anomalies matrix and $Z_{k/k}$ is the update anomalies matrix. In other words, the prediction estimate $\overline{x}_{k/k-1}$ (respectively the update estimate $\overline{x}_{k/k}$) of the turbulent phase is given by the mean of the $m$ members of the ensemble $\mathcal{X}_{k/k-1}$ (respectively $\mathcal{X}_{k/k}$). All these members will constitute together a cloud of points in the state space and the spreading of this ensemble characterizes the prediction error variances by $ \Sigma_{k/k-1} = Z_{k/k-1}\times Z^{\rm T}_{k/k-1}$ (respectively the update error variances by $\Sigma_{k/k} = Z_{k/k}\times Z^{\rm T}_{k/k}$).\\
\indent However, in this EnKF formulation, it has been shown the need to add in the measurements $y_{k}$ some perturbations (random variables calculated from a distribution with a covariance matrice equal to $\Sigma_{w}$: for more details see \cite{Burgers-a-98}), but this use of 'pertubated measurements' introduces sampling errors that reduce the update covariance matrix accuracy.\\
\indent Another variant has been developed in order to form the update ensemble $\mathcal{X}_{k/k}$ deterministically, avoiding part of the sampling noise. In particular, the update estimation is not given by the mean $\overline{x}_{k/k}$ of the $m$ members (model states) of this ensemble $\mathcal{X}_{k/k}$, but directly by the estimate $\hat{x}_{k/k}$ from the KF Eq. \eqref{Est_maj}. In the following, we present this deterministic algorithm for transforming the prediction ensemble $\mathcal{X}_{k/k-1}$ into an update ensemble $\mathcal{X}_{k/k}$: it is called the Ensemble Transform Kalman Filter (ETKF).\\
\indent For the initial prediction ensemble, one can simply use $\mathcal{X}_{1/0} = \{x^{1}_{1/0};...;x^{m}_{1/0}\} = \{0;...;0\}$.

\subsubsection{The update step}

The ETKF-based method doesn't explicitly calculate the empirical covariance matrices, but transforms prediction anomalies $Z_{k/k-1}$ into update anomalies $Z_{k/k}$ by an Ensemble Transform Matrix (ETM) $T_{k}$ with the relation: 
\begin{equation}\label{Cov_emp_T}
Z_{k/k} = Z_{k/k-1}\times T_{k},
\end{equation}%
such that the empirical prediction covariance matrix $\Sigma_{k/k-1}$ and the empirical update covariance matrix $\Sigma_{k/k}$, defined by Eq. \eqref{Cov_emp}, both match the theoretical Eqs. \eqref{Cov_maj} and \eqref{K_Gain}. There are different solutions for the ETM $T_{k}$ and following the mathematical approach in \cite{Bishop-a-01,Tippett-a-03,Ott-a-04,Sakov-a-08}, a general form that satisfies Eqs. \eqref{Cov_maj} and \eqref{K_Gain}, and Eqs.\eqref{Cov_emp} and \eqref{Cov_emp_T}, is:
\begin{equation}\label{Mat_T}
T_{k} = [{\rm Id} + ({\rm C_{1}}Z_{k/k-1})^{\rm T}\times\Sigma_{w}^{-1}\times({\rm C_{1}}Z_{k/k-1})]^{-1/2}.
\end{equation}
\indent We emphasize that, in this study, we assume the measurement noise covariance matrix $\Sigma_{w}$ to be a stricly positive diagonal matrix (no correlation between different subapertures). Given the Eigen Value Decomposition (EVD) of the matrix ${\rm Id} + ({\rm C_{1}}Z_{k/k-1})^{\rm T}\Sigma_{w}^{-1}({\rm C_{1}}Z_{k/k-1})$ whose size is $m\times m$, the solution for the ETM $T_{k}$ can be obtained with this relation:
\begin{equation}\label{Mat_T_QG}
T_{k} = Q_{k}\times\Gamma_{k}^{-1/2}\times Q_{k}^{\rm T},
\end{equation}
where the orthogonal matrix $Q_{k}$ contains the normalized eigenvectors of the $m\times m$ matrix in the square brackets of Eq. \eqref{Mat_T} and the diagonal matrix $\Gamma_{k}$ contains the eigenvalues of the EVD. Since ${\rm Id} + ({\rm C_{1}}Z_{k/k-1})^{\rm T}\Sigma_{w}^{-1}({\rm C_{1}}Z_{k/k-1})$ is a positive definite real symmetric matrix, its eigenvalues are real, strictly positive and its eigenvectors are orthogonal.\\
\indent Both prediction and update covariance matrices belong to the same linear subspace spanned by the ensemble prediction anomalies $Z_{k/k-1}$. A distinguishing feature of the update anomalies $Z_{k/k}$ produced by the ETKF is that they are orthogonal under the inner product (defining also a Euclidian norm): ${\rm <z_{1} | z_{2}> = z_{1}^{\rm T}C_{1}^{\rm T}\Sigma_{w}^{-1}C_{1}z_{2}}$. However, we have to notice that no more than $m-1$ independent update anomalies are generated from the expression in Eq. \eqref{Cov_emp} because the sum of the $m$ prediction anomalies (columns of $Z_{k/k-1}$) is equal to zero: therefore, the rank of the empirical covariance matrices $\Sigma_{k/k-1}$ and $\Sigma_{k/k}$ is equal to at most $m-1$.\\
\indent Actually, in this update scheme with the ETKF-based control law, the Eq. \eqref{Est_maj} of the update estimate $\hat{x}_{k/k}$ with the Kalman gain $H_{k}$ given by Eq. \eqref{K_Gain}, is completely transformed by using the anomalies matrix $Z_{k/k-1}$ and the ETM $T_{k}$ with the Sherman-Morrison-Woodbury identity (see Appendix A). As $\Sigma_{w}$ is a strictly positive diagonal matrix, it is straightfoward to calculate $\Sigma_{w}^{-1/2}$. Let us define the vector $S_{inov}= \Sigma_{w}^{-1/2}(y_{k} - \overline{y}_{k/k-1})$ where $\overline{y}_{k/k-1} = {\rm C_{1}}\times\overline{x}_{k/k-1} - {\rm D_{1}N}\times u_{k-2}$, and the matrix $S_{cz} = \Sigma_{w}^{-1/2}{\rm C_{1}}Z_{k/k-1}$, the expression for the update estimate is therefore:
\begin{equation}
\hat{x}_{k/k} = \overline{x}_{k/k-1} + Z_{k/k-1}S_{cz}^{\rm T}(S_{inov}-S_{cz}Q_{k}\Gamma_{k}^{-1}Q_{k}^{\rm T}S_{cz}^{\rm T}S_{inov}).
\end{equation}
The matrix of the $m$ members of the update ensemble $\mathcal{X}_{k/k}$ is obtained with:
\begin{equation}
X_{k/k} = \sqrt{m-1}\times Z_{k/k} + [\hat{x}_{k/k};...;\hat{x}_{k/k}].
\end{equation}

\subsubsection{The prediction step}

In order to obtain each of the $m$ members of the prediction ensemble, we have to compute:
\begin{equation}\label{Mem_pre}
x_{k+1/k}^{i} = {\rm A_{1}}\times x_{k/k}^{i} + {v}_{k+1}^{i} \qquad (\rm for \qquad 1 \leq i \leq m),
\end{equation}  
where ${v}_{k+1}^{i}$ is a zero-mean random Gaussian vector characterising the model noise (with a covariance matrice equal to $\Sigma_{v}$). The prediction estimate $\overline{x}_{k+1/k}$ is given by the mean of the $m$ members of the prediction ensemble $\mathcal{X}_{k+1/k}$ and the prediction anomalies matrix $Z_{k+1/k}$ is still given by the expression of Eq. \eqref{Cov_emp}.

\subsection{Theoretical numerical complexity with non-stationary models}

Let us note $n$, the dimension of the state vector $x_{k}$ and $p$, the dimension of the observation vector $y_{k}$ in Eq. \eqref{Est_maj}. In order to determine the theoretical numerical cost with a \textit{non-stationary} turbulence model, we have to calculate the total number of all the MVM computed during the update step and the prediction step. Actually, the significant computational cost of the ETKF method is the update step. It can be proved (Appendix B) that the resultant theoretical cost is:
\begin{equation}
O(m^{3} + m^{2}\times(n + p)),
\end{equation}
which is therefore linear over $n$ and $p$ with a proportional factor equal to $m^{2}$. It will actually depend on the relative magnitudes of the parameters $m$, $n$ and $p$. With this ETKF-based method, it is numerically more efficient if the value of $m$ remains smaller than the values of $n$ and $p$.

\subsection{ETKF limitations for AO systems on ELTs}

In the ETKF-based method, there are two related limitations \cite{Bocquet-a-11} by using an ensemble with $m$ members for the calculations of empirical covariance matrices with Eqs. \eqref{Cov_emp} and \eqref{Cov_emp_T}.\\
\indent The first limitation is the model space dimension of the finite size ensemble $\mathcal{X}$ much lower than the one on the whole pupil of the telescope. If the ensemble $\mathcal{X}$ has $m$ members, then the empirical prediction covariance matrices $\Sigma_{k/k-1}$ describe uncertainty only in the $(m-1)$-dimensional subspace spanned by the ensemble. The global update will allow adjustments to the system state only in this subspace which is usually rather small compared to the total dimension of the model space on the whole pupil of the telescope. As a small ensemble has few degrees of freedom available to represent estimation errors, this situation leads to sampling errors and a loss of accuracy with an underestimation of the true covariance matrices. Thus, these empirical covariance matrices calculated from the $m$ members will not be able to match the model rank (the effective number of degrees of freedom of the model) unless this number of members increases considerably: this leads to an extremely high theoretical numerical complexity and no possibility of a realistic implementation on ELTs. The idea is therefore to perform \textit{locally} all the updates so that, with different linear combinations of the ensemble members in various domains, the global update explores a much higher dimensional space.\\
\indent The second (though related) limitation is the spurious correlations over long spatial distances produced by a limited size ensemble $\mathcal{X}$. The empirical covariance matrices are calculated with the statistical estimator (Eq. \eqref{Cov_emp}) where the ensemble $\mathcal{X}$ of model states is considered as a statistical ensemble and the models (turbulence, WFS ...) are imperfect. Therefore, the correlation calculations from the ensemble sample assign non-zero values to correlations between variables separated by a large distance (compared to the Fried parameter $r_{0}$), which leads again to an underestimation of the true covariance matrices. It can be shown that variances of these spurious random correlations decrease when the number of members increases: however, it is not suitable for an implementation on ELTs. The idea is to determine the system's characteristic correlation distance, and then perform again \textit{locally} so that the \textit{local update} should ignore ensemble correlations for distances larger than this correlation distance/length (which is not equal to $r_{0}$).


\section{The Local ETKF : an intrinsically parallel algorithm}

In the previous section, we have described two related shortcomings with the ETKF-based control law, which lead to underestimate the theoretical covariance matrices. In order to overcome these drawbacks, we have proposed the necessity to use local updates. Thus, a new version called the Local ETKF \cite{Ott-a-04,Hunt-a-07} has been developed using domain decomposition and localizations. There are two common localization methods in the EnKF-based approaches. The first one is the Local Analysis (LA): the local update is performed explicitly by considering only the measurements from a local region surrounding the local domain by building a virtual local spatial window. This is equivalent to setting ensemble anomalies outside a local window to zero during the update. The second one is the Covariance Localization (CL) where the local update is performed implicitly: the prediction covariance matrices are tapered by a Schur-product with a distance-based correlation matrix. It increases the rank of the modified covariance matrix and masks spurious correlations between distant state vector elements. As the two methods yield very similar results \cite{Sakov-a-10a}, we present in this section the LA: thus, the idea of the Local ETKF is to split up the pupil of the telescope into various local domains on which all calculations of the update step are performed independently.

\subsection{Description of the update step}

In order to understand the principles of the Local ETKF, let us take the following example with a 16 m diameter telescope: Fig. 3 shows the locations of the valid actuators (blue dots) on the pupil sampled by a $32\times32$ SH-WFS with a Fried geometry: therefore, each square area between 4 valid actuators represents a subaperture. In the following, we describe the Local ETKF-based control law on a \textit{zonal} basis: the turbulent phase is estimated \textit{on the locations} of all DM's actuators.
\begin{figure}[htbp]
\centering\includegraphics[width=3cm]{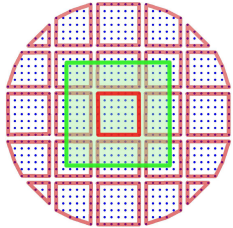}
\caption{Partition of the actuators domains on a 16 m telescope pupil with a 32$\times$32 SH-WFS.}
\end{figure}
In this example, the domain decomposition is made of 25 (red) estimation subpartitions, called the \textit{local domains}: each of them is composed by a fixed number of actuators. During the update step, for each local domain, a local update estimate (made up of all estimations on the actuators' locations in this local domain) is performed locally from the measurements coming only from the connected \textit{observation region} (made up of all subapertures \textit{including} and  \textit{surrounding} the local domain). For instance, the observation region (measurements region) connected to the red central local domain (estimation domain) is the green one.\\
\indent With the Local ETKF-based method, only the update step will change: on \textit{each local domain}, there is an \textit{independent} update. Therefore, this update step scheme enables highly parallel computations of markedly less data.\\
\indent We emphasize that it is essential to avoid as far as possible discontinuities between two neighboring local updates coming from two different local domains. To be precise, the two results of two update estimates coming from two nearby actuators on the border of two different local domains have to be similar. This can be ensured by choosing similar sets of observations for neighboring actuators, i.e. by taking two overlapping observation sets coming from two different neighboring local regions. Moreover it is necessary to have a smoothed localization by gradually increasing the uncertainty assigned to the observations until beyond a certain distance they have infinite uncertainty and therefore no influence. This can be done by multiplying (in Eq. \eqref{Sicz_majLoc} hereafter) the square root inverse of the measurement noise covariance matrix by a decreasing function (from one to zero) as the distance of the observations from the center of the local observation region increases \cite{Hunt-a-07,Sakov-a-10a}.

\subsection{Basic mathematical formulation of the Local ETKF}

We introduce a new notation where a tilde denotes the local version of the corresponding variable. For example: $\widetilde{Z}_{k/k-1}$ means the \textit{local} prediction anomalies used for a given \textit{local} domain d; $\widetilde{y}_{k}$ means the \textit{local} measurements from the \textit{local} observation region including and surrounding a given \textit{local} domain d. In order to facilitate the understanding, we give also the size of each local variable obtained in the calculations. Therefore, we use the new variables $n_{d}$ and $p_{d}$, which are defined as $n$ and $p$: they are respectively the dimension of the local state vector and the dimension of the corresponding local measurement vector \textit{on a local domain d}.\\
\indent Thus, for the update step, on \textit{each local domain d}, we have to determine \textit{independently} each local update $\widetilde{\hat{x}}_{k/k}$ by computing:
\begin{equation}
\widetilde{\overline{x}}_{k/k-1} \qquad (n_{d},1) \qquad {\rm and} \qquad \widetilde{Z}_{k/k-1} \qquad (n_{d},m)
\end{equation}
\begin{equation}\label{Sicz_majLoc}
\widetilde{S}_{inov}= \widetilde{\Sigma}{_{w}^{-1/2}}(\widetilde{y}_{k} - \widetilde{\overline{y}}_{k/k-1})  \qquad (p_{d},1) \qquad {\rm and} \qquad \widetilde{S}_{cz} = \widetilde{\Sigma}{_{w}^{-1/2}}\widetilde{\rm C}_{1}{Z}_{k/k-1} \qquad (p_{d},m)
\end{equation}
\begin{equation}
{\rm EVD \,\,\,\, of}\,\,\,\,({\rm Id} + \widetilde{S}{_{cz}^{\rm T}}\times\widetilde{S}_{cz})\qquad\longrightarrow\qquad\widetilde{Q}_{k} \qquad (m,m) \qquad{\rm and} \qquad \widetilde{\Gamma}_{k} \qquad (m,m)
\end{equation}
\begin{equation}\label{Est_majLoc}
\widetilde{\hat{x}}_{k/k} = \widetilde{\overline{x}}_{k/k-1} + \widetilde{Z}_{k/k-1}\widetilde{S}{_{cz}^{\rm T}}(\widetilde{S}_{inov} - \widetilde{S}_{cz}\times\widetilde{Q}_{k}\widetilde{\Gamma}{_{k}^{-1}}\widetilde{Q}{_{k}^{\rm T}}\times\widetilde{S}{_{cz}^{\rm T}}\widetilde{S}_{inov}) \qquad (n_{d},1)
\end{equation}
\begin{equation}
\widetilde{X}_{k/k} = \sqrt{m-1}\times\widetilde{Z}_{k/k-1}\times\widetilde{Q}_{k}\widetilde{\Gamma}{_{k}^{-1/2}}\widetilde{Q}{_{k}^{\rm T}} + [\widetilde{\hat{x}}_{k/k};...;\widetilde{\hat{x}}_{k/k}] \qquad (n_{d},m).
\end{equation}
With the concatenation of all those small matrices $\widetilde{X}_{k/k}$, we can finally compute the update members matrix ${X}_{k/k}$.\\ 
\indent Then, the prediction step is computed \textit{globally} on the whole pupil (as it is done with the ETKF in section 3.1.2): the computation of the $m$ predicition members (Eq. \eqref{Mem_pre}) for the prediction matrix ${X}_{k+1/k}$, the prediction estimate $\overline{x}_{k+1/k}$ and the prediction anomalies matrix $Z_{k+1/k}$.\\
\indent For the sake of understanding the ability of this algorithm to handle non-stationary behaviors, each local update estimate (Eq. \eqref{Est_majLoc}) can be rewritten by factoring $\widetilde{S}_{inov}$ (to the right):
\begin{equation}
\widetilde{\hat{x}}_{k/k} = \widetilde{\overline{x}}_{k/k-1} + \widetilde{H}_{k}\times(\widetilde{y}_{k} - \widetilde{\overline{y}}_{k/k-1})
\,\,\,\,\,\,\,\, {\rm with} \,\,\,\,\,\,\,\, \widetilde{H}_{k} = \widetilde{Z}_{k/k-1}\widetilde{S}{_{cz}^{\rm T}}(I - \widetilde{S}_{cz}\widetilde{Q}_{k}\widetilde{\Gamma}{_{k}^{-1}}\widetilde{Q}{_{k}^{\rm T}}\widetilde{S}{_{cz}^{\rm T}})\widetilde{\Sigma}{_{w}^{-1/2}}
\end{equation}
where $\widetilde{H}_{k}$ is called the local Kalman gain on the local domain d. With this last expression, we can clearly see that all local Kalman gains are computed during \textit{each} update step, without the need to resolve a DARE. Therefore, we can have a non-stationary command with small modifications of the matrices $\widetilde{Z}_{k/k-1}$ and $\Sigma_{w}$: there will be far fewer transitional issues than what happens with the LQG command and the full recomputation of all control matrices. Moreover, during the prediction step, by using a (to be defined) identification procedure, we can update the AR model (matrices $A_{1}$ and $\Sigma_{v}$) and take into account this modification with Eq. \eqref{Mem_pre}.

\subsection{Theoretical numerical complexity with non-stationarity models}

Let us note $n_{max}$ the largest value of the dimensions $n_d$ of all local update estimates on all local domains, and $p_{max}$ the largest value of the dimensions $p_d$ of all observation vectors on all observation regions. With the Local ETKF, we have to distinguish two computational costs.\\
The theoretical cost of the update step \textit{on each local domain} is: $O(m^{3} + m^{2}\times(n_{max} + p_{max}))$.\\
This complexity is again linear over $n_{max}$ and $p_{max}$ with a proportional factor equal to $m^{2}$. We emphasize that the values of $n_{max}$ and $p_{max}$ can be significantly smaller than $n$ and $p$ when the pupil of the telescope has been split up into many subpartitions: therefore, by using the result given in Appendix B.1, the total number of operations on each local domain can be much smaller. Moreover, it will only depend on the size of the subpartitions (number of actuators) and not on the diameter of the telescope (see table 1 in section 5.3). Actually this total number of operations will depend on the relative magnitudes of $m$, $n_{max}$ and $p_{max}$ which are determined by an optimal trade-off between the size of the local domains, the size of the observation regions and the expected image quality (see the various simulations in section 5.2).\\
For the prediction step, the theoretical cost is: $O(m\times n)$.\\
\indent A key point is that both the calculations during the update step and the prediction step can be easely parallelized, which considerably speeds up the algorithm for AO systems on ELTs (see discussion in section 5.3).

\subsection{Local phase estimation and differential pistons issue}

One particular consequence of the Local ETKF approach is the differential pistons issue. Indeed, as described in section 4.1, we perform a partition of the pupil of the telescope into various local domains, on which one local update estimate of the turbulence phase will be produced. Afterwards, one global update estimate (for each member of the ensemble) and only one global prediction estimate are performed on the \textit{whole} pupil. Turbulence continuity from a local domain to its neighbors, is ensured to some extent in the overlapping of the observation regions, that includes for each connected local domain, the surrounding subapertures. However this continuity does not extend to piston which is not measured by the WFS.\\
\indent In other words, turbulence estimation on local domains is piston-free, which means \textit{ipso facto} that a piston continuity issue will arise when combining all local update estimates of the turbulent phase. In order to calculate these differential pistons and to remove them, we have now implemented a first effective and fast algorithm inside the AO loop, using a least-squares based method, which globally minimizes the local discontinuities and ensure phase continuity on the whole pupil.\\
\indent Though straightforward, this algorithm allows improving significantly the Local ETKF performance (see section 5.1). However, this algorithm does not yet take into account the noise propagation at the level of turbulence phase estimation, and thus the influences on local pistons. Of course, this is a foreseen improvement, that could simply rely on a regularized least-square method, taking advantage of our knowledge of phase estimation accuracy, embedded in the empirical update estimation error covariance matrix given by $\Sigma_{k/k}$.


\section{First simulation results with a classical AO system (D = 16 m)}

In order to validate some of the items discussed in sections 3 and 4, we have implemented the three previous control laws (based on the KF, the ETKF and the Local ETKF) under the OOMAO Matlab environment \cite{Conan-p-14} and we have also developed our OpenMP parallelized version. For this Single Conjugate AO (SCAO) system simulation, we consider a 16 m diameter telescope (without central obscuration) with a 32$\times$32 microlenses S-H WFS. Only the subapertures with a surface enlightened more than 50 \% are considered  and the number of valid subapertures is 812 that means $p = 1624$. The linear response of the S-H WFS is emulated by a matrix which calculates differences of the phase in two directions at the edges of each subaperture \cite{Petit-a-09,Massioni-a-11}. The AO system works in a closed-loop at 500 Hz. There is a two-frame delay between measurement and correction. We assume that the DM has an instantaneous response and the coupling factor of the actuators is 0.3. Using a zonal basis, the phase is estimated only on the actuators' locations and the number of valid actuators is 877 that means $n = 1754$.\\
\indent The criterion used in the comparisons is the coherent energy, defined by $E_{coh} = \exp(-\overline{\sigma^{2}_{res}})$, where $\overline{\sigma^{2}_{res}}$ is the temporal mean of the spatial variances of the residual phase $\varphi^{res}$.\\
\indent The loss of performance (in \%) with the Local ETKF is calculated by using the optimal solution given by the KF:
\begin{equation}
loss = (E_{coh}^{KF} - E_{coh}^{Local ETKF})/E_{coh}^{KF}\times 100.
\end{equation}
\indent For the simulation of the atmosphere, we consider a Von Karman turbulence with a stationary model: $r_{0}$ \,=\, 0.525 m (at 1.654 $\mu$m), $L_{0}$ \,=\, 25 m, $\lambda$ \,=\, 1.654 $\mu$m (for both WFS's and observation's wavelengths). Using Taylor's hypothesis, we can generate a superimposition of 3 turbulent phase screen layers moving at 7.5 $ms^{-1}$, 12.5 $ms^{-1}$ and 15 $ms^{-1}$, with a relative $C_{n}^{2}$ profile of 0.5, 0.17 and 0.33 respectively. Phase screens are generated on a 320$\times$320 grid, with 10$\times$10 points per each subaperture: the correction phase is therefore obtained by multiplying the prediction estimate on the actuators' locations with an influence matrix of the DM.\\
\indent For the turbulence temporal model in the KF, in the ETKF and in the Local ETKF-based control laws, we have chosen a \textit{first} order Auto-Regressive model (AR1) \cite{Kulcsar-a-06,Petit-a-09}. Each value of the coherent energy has been calculated with several simulations of 5000 iterations (10 sec).

\subsection{Convergences of the ETKF and the Local ETKF to the KF}

The mathematical convergence of the Ensemble Kalman Filters to the Kalman Filter in the limit for large ensembles has been proved in \cite{LeGland-b-11} for linear forecast models.
\begin{figure}[H]
\centering{\rotatebox{270}{\includegraphics[width=5cm]{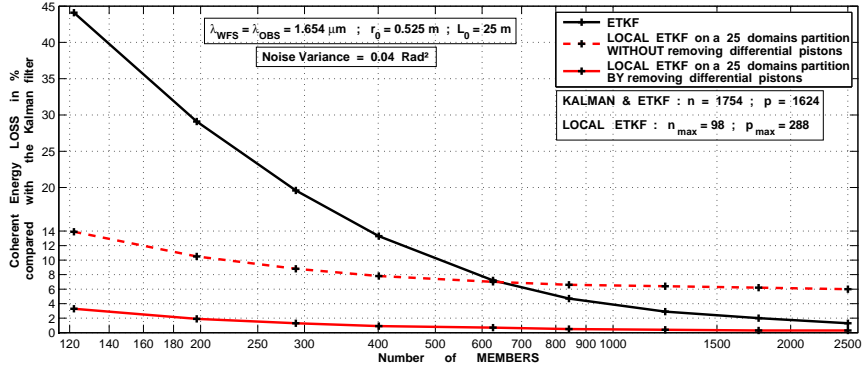}}}
\caption{Convergences of the ETKF and the Local ETKF performances to the KF.}
\end{figure}
Therefore, the first step is to obtain this theoretical result with the simulations for the ETKF: it was done with one given value of noise variance equal to 0.04 $\rm Rad^{2}$ for which the KF gives a coherent energy equal to 88.5 \%. In Fig. 4, each curve gives the coherent energy loss in \% between the performance obtained with a control law (ETKF or Local ETKF on a 25 domains partition \textit{without} removing or \textit{by} removing differential pistons) and the one given by the KF. This loss depends on the number of members and decreases to zero when the value of $m$ increases. The performance provided by the Local ETKF \textit{without removing} differential pistons (red dashed curve) is better than the one provided by the ETKF until a given number of members ($\simeq 626$). Indeed, concerning the phase estimation with the Local ETKF, there is a problem of reattachment between the various local domains discussed in section 4.4: as the local update estimates of the turbulent phase are done separately on each local domain, we have to take into account the differential pistons and to reconstruct the entire phase estimate on the whole pupil of the telescope with all these various small turbulent phase estimates.\\
\indent As the red solid curve shows, once differential piston has been removed with our first least-squares based method, the coherent energy loss compared with the KF is always lower than 4 \% for numbers of members greater than 100. With only 122 member, this loss is already about 3.5 \% on a 25 domains partition (see also Fig. 5).

\subsection{Performance of the Local ETKF with different partitions of the pupil of the telescope}

The second step is to study different kinds of partitions and their influences on the performances for a fixed value of noise variance. Figure 5 shows the coherent energy losses (compared with the KF) obtained with the Local ETKF in the case of five kinds of partitions and various values of members in the ensemble.
\begin{figure}[htbp]
\centering{\rotatebox{270}{\includegraphics[width=6cm]{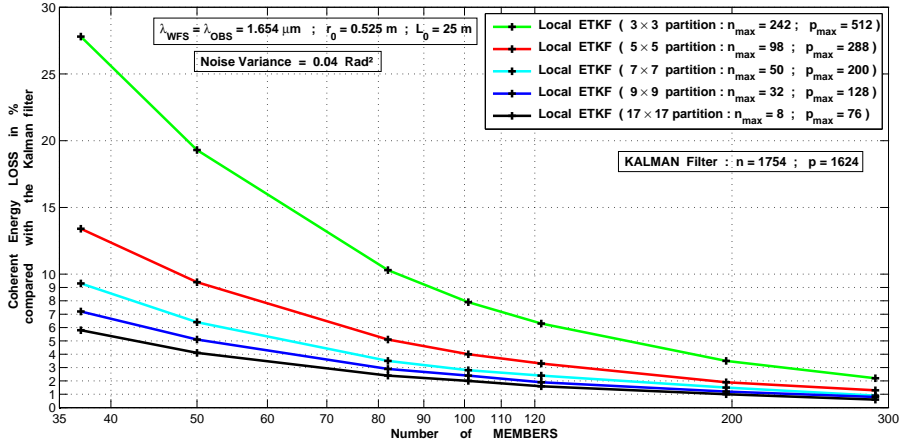}}}
\caption{Coherent energy loss with different partitions and a fixed noise variance.}
\end{figure}
As already shown in Fig. 4, for a given partition, the larger is the number of members, the smaller is the loss of performance. In the same way, \textit{for a fixed value of members}, the larger is the number of subpartitions (or the smaller is the number of actuators per local domain), the smaller is the loss of performance: it confirms the specific problems of the ETKF explained in section 3.3 and the solution given by the Local ETKF. The localization improves the efficiency of the ETKF-based approach because, firstly the ensemble needs only to encompass the uncertainty within each of the small local domains (with small numbers of degrees of freedom), and secondly the spurious long-range correlations produced by a limited ensemble size are removed. What is also important in this simulation with a 16 m telescope, is that we can choose a partition of the pupil, for which the loss of performance is less than 3\%: for example with the $9\times9$ partition and a number of members equal to 101 (it will be also true with a larger number of subpartitions or a larger number of members).\\
\indent The third step is to study the influence on the performance for different values of noise variance with a fixed number of members (m = 101).
\begin{figure}[htbp]
\centering{\rotatebox{270}{\includegraphics[width=6.2cm]{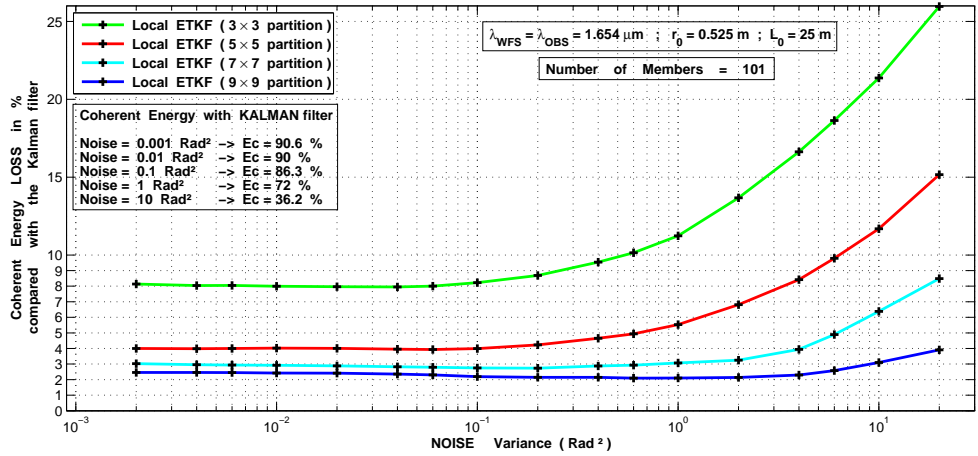}}}      
\caption{Coherent energy loss with different partitions and 101 members.}
\end{figure}
Let us recall that in order to remove the differential pistons, our first least-squares based method doesn't take into account the measurement error covariance. Of course, for a given partition, when the noise variance increases, the loss of performance increases too. Nevertheless, as Fig. 6 shows, the larger is the number of subpartitions, the smaller is the loss of performance when the noise variance increases. Moreover, for the $9\times9$ partition (with a maximum of 16 actuators per local domain), the rosbustness of the Local ETKF compared to the KF seems to be very good for a large range of noise variance, while the correction of the differential pistons can be still improved.\\
\indent Compared performance along the time for the KF and the Local ETKF is proposed in Fig. 7 for various kinds of partitions. This figure shows that convergence speed and stability of the Local ETKF is very similar to the KF. 
\begin{figure}[htbp]
\centering{\rotatebox{270}{\includegraphics[width=6cm]{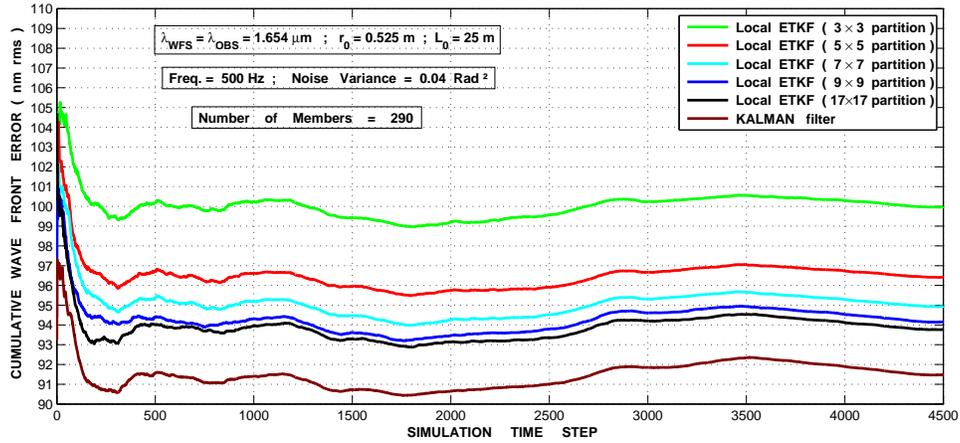}}}
\caption{Cumulative Wave Front Error (in nm rms) with 290 members.}
\end{figure}
It underlines once more that increasing the number of subpartitions allows to reduce the loss of performance compared with the KF and that a very small difference in performance is brought going from a $9\times9$ to a $17\times17$ partitions.

\subsection{First speed-tests for a parallel implementation}

The complexity given in section 4.3 is only theoretical. We want to evaluate the performance of a real parallel implementation of this intrinsically parallel algorithm based on the Local ETKF: therefore, we made some runtime tests in order to demonstrate that this method could be used for a 40 m telescope with a frequency up to 500 Hz using nowadays computers.\\
\indent Let us consider the 16 m telescope for which the loss of coherent energy compared with the KF is less than 3 \% with the $9\times9$ partition (total of 69 domains) and 101 members. In the case of a 40 m telescope with a $80\times80$ SH-WFS (and the same $r_{0}$), by keeping local domains with the same maximum number of actuators, we can reach a similar coherent energy loss with the same number of members (m = 101). For the 16 m telescope with the $9\times9$ partition, there is no more than 16 actuators per domain. For the 40 m telescope, in order to have no more than 16 actuators per domain, we must take a $21\times21$ partition (total of 373 domains). In this AO configuration, our numerical simulations have confirmed that the loss of coherent energy is indeed less than 3\%. Each cycle of the AO loop has 3 steps: all the local updates (calculated independently on each local domain), the prediction and the calculation of voltages $u_k$.\\
\indent Therefore, we propose to consider a multi-core architecture (single computer or computation cluster), where each core is assigned to only one local domain and performs the local update computations. By using the results of Table 3 and Table 4 in Appendix B, Table 1 gives the theoretical numbers of floating-point operations during one cycle of the AO loop.
\begin{table}[H]
\centering\caption{Numbers of operations with m = 101 during one cycle of the AO loop.}
\begin{tabular}{|c|c|c|c|c|c|c|c|c|}
\hline
$D$   & $n$ & $p$ & Partition & $n_{max}$ & $p_{max}$ & Update & Prediction & $u_{k}$\\
(m)   &     &     &           &           &           & (1 domain) &        &        \\
\hline
 16 & 1754 & 1624 & $9\times9$ & 32 & 128 & $7.65\times10^{6}$ & $0.71\times10^{6}$ & $0.76\times10^{6}$\\
 40 & 10370 & 10048 & $21\times21$ & 32 & 128 & $7.65\times10^{6}$ & $4.19\times10^{6}$ & $6.56\times10^{6}$\\
\hline
\end{tabular}
\end{table}
\indent Table 2 gives real speed-tests with a first version of our OpenMP parallelized code: we used a workstation with two Intel(R) Xeon(R) E5-2680 v2 CPUs (total of 20 cores). We have measured the time required for calculating, the update step for one local domain by using only one core, the prediction step by using all the 20 cores and the voltages $u_k$ by using also all the 20 cores.
\begin{table}[H]
\centering\caption{Runtime (in msec) with m = 101 during one cycle of the AO loop.}
\begin{tabular}{|c|c|c|c|c|c|c|}
\hline
$D$ & Partition & Update for 1 domain & Prediction        &  $u_{k}$          & Total\\
(m) &           & (using 1 core)      & (using 20 cores)  &  (using 20 cores) &      \\
\hline
16 & $9\times9$ & 0.94 & 0.14 & 0.1 & 1.18 \\
40 & $21\times21$ & 0.94 & 0.7 & 1 & 2.64 \\
\hline
\end{tabular}
\end{table}
\indent In the AO configuration on a 16 m telescope, we have only 69 domains. Nowadays we can have a single computer with 120 cores (8$\times$15-cores CPUs) similar to the cores in our test workstation. Table 2 shows that, with this kind of single computer (which has more cores than the number of requested domains), our algorithm needs a total of 1.18 ms for one cycle of the AO loop. Therefore, we can easely implement and use our algorithm with a 500 Hz frequency for the AO loop. For the AO configuration on a 40 m telescope, the situation is not basically different: there are many more local domains, but during the update step, computations on each local domain take exactly the same time as in the 16 m case. The time required for the prediction step and the $u_k$ calculation became not negligible: however both of these computations (prediction estimate $\overline{x}_{k+1/k}$ and voltages $u_k$) are rather well scalable with a larger number of cores. Right now, we cannot have a single computer with so many cores (373), and we have to consider computation cluster which implies additional time for communication stage between the update step and the prediction step. On our computation cluster with infiniBand 4x QDR network, this communication stage takes $\sim 2$ msec. However the last generation of infiniBand (12x EDR) is claimed to be almost 10 times faster, which makes times for communication reasonable small and a real implementation on a 40 m telescope achievable.


\section{Conclusion}

We have presented a new control law, the Local ETKF, based on the LQG approach, a KF adaptation with localizations for large-scale AO systems and the assumption of a perfect DM dynamics. The advantages of this proposed method are significant. Firstly, as all the local Kalman gains can be calculated with empirical covariance matrices at each cycle of the AO loop, it enables to deal with non-stationary behaviors (turbulence, vibrations). Secondly, as the structure of this algorithm is intrinsically parallel, its implementation on ELTs can easily be done on a parallel architecture (CPUs/GPUs cluster) with a reduced computational cost: let us remind that the complexity of the update step does not depend on the diameter of the telescope but only on the maximum number of actuators per each local domain, which significantly speeds up the algorithm. The more subpartitions in the pupil of the telescope, the better is the performance, both in terms of coherent energy and of runtime. In our simulations with a Von Karman turbulence and a SCAO system with an adequate partition of the pupil, the loss of coherent energy compared with the optimal solution given by the KF can be less than 3 \%. We have presented numerical simulations in the case of a SCAO system with an AR1 turbulence model on a zonal basis, but we can already extend this method with an AR2 model. For the runtime tests we have already developed an OpenMP parallelized version, but we are also currently working on a new version with GPUs on the COMPASS platform \cite{Gratadour-p-14}.\\
\indent In the short-term, there are different points to study more in details. The first one is the influence of the turbulence charateristics on the size of the local observation regions which reflects the distance, called the localization length, over which the correlations calculated with the ensemble's members are not meaningful. In the same way, we need to evaluate the influence of the spider arms on the choice of the subpartitions on the pupil of the telescope. In order to resolve the problem of differential pistons, we have to improve our fast least-squares based method by taking into account the error covariance matrices. Then, the robustness of the Local ETKF must be also studied when the parameters of the turbulence change. We must indeed estimate the impact of the turbulence model error (wind speed, turbulence profile, $L_{0}$), first in absence of turbulence model update. Then, assuming some turbulence model identification and update, one shall consider the gain brought by this non-stationary control solution, its stability and the speed of convergence.\\
\indent In the long-term, we have of course to demonstrate the potentials of the Local ETKF in the case of wide field tomographic AO and to consider the extension to limited DM dynamics. Afterwards, two other important aspects already developed in geophysics must be also explored in AO. The first one is when the operator C in Eq. \eqref{Syst_mod} is non-linear \cite{Bocquet-a-12}. The EnKF-based method does not require a linearized model, an advantage over the KF-based method, and this could be very suitable for non-linear WFS. The second one is the possibility of asynchronous observations assimilation \cite{Hunt-a-07,Sakov-a-10b}, for the multi-rate case in the prospect of Natural Guide Star and Laser Guide Star wavefront sensing for wide field tomographic AO.


\section*{Appendix A: Mathematical expression of the update estimate $\hat{x}_{k/k}$ for the ETKF}

For computational reasons, it is better to change the original expression of the update estimate $\hat{x}_{k/k}$ in Eq. \eqref{Est_maj} by using the following version of the Sherman-Morrison-Woodbury identity:
\begin{equation}\label{Id_SMW}
(U\times U^{\rm T} + S\times S^{\rm T})^{-1} = (S^{-1})^{\rm T}\{S^{-1} - (S^{-1}U)[{\rm Id} + (S^{-1}U)^{\rm T}(S^{-1}U)]^{-1}(S^{-1}U)^{\rm T}S^{-1}\}.
\end{equation}
\indent By replacing the two expressions from Eqs. \eqref{K_Gain} and \eqref{Cov_emp} in Eq. \eqref{Est_maj}, we obtain:
\begin{equation}\label{Est_majA1}
\hat{x}_{k/k} = \overline{x}_{k/k-1} + Z_{k/k-1}Z_{k/k-1}^{\rm T}{\rm C_{1}^{T}}({\rm C_{1}}Z_{k/k-1}Z_{k/k-1}^{\rm T}{\rm C_{1}^{T}} + \Sigma_{w})^{-1}(y_{k} - \overline{y}_{k/k-1}).
\end{equation}
\indent We can note $U = {\rm C_{1}}Z_{k/k-1}$ and as $\Sigma_{w}$ is a strictly positive diagonal matrix, we can note $S = \Sigma_{w}^{1/2}$ which is very easy to compute and to invert. By using the notations U and S, we can identify $({\rm C_{1}}Z_{k/k-1}Z_{k/k-1}^{\rm T}{\rm C_{1}^{T}} + \Sigma_{w})^{-1}$ as the left-hand side of Eq. \eqref{Id_SMW} and by using the right-hand side of Eq. \eqref{Id_SMW} in Eq. \eqref{Est_majA1}, we can obtain a new expression for $\hat{x}_{k/k}$:
\begin{equation}\label{Est_majA2}
\begin{aligned}
\hat{x}_{k/k} & = \overline{x}_{k/k-1} + Z_{k/k-1}(\Sigma_{w}^{-1/2}{\rm C_{1}}Z_{k/k-1})^{\rm T}\{\Sigma_{w}^{-1/2}(y_{k} - \overline{y}_{k/k-1})-\Sigma_{w}^{-1/2}{\rm C_{1}}Z_{k/k-1}\times\\
&[{\rm Id} + (\Sigma_{w}^{-1/2}{\rm C_{1}}Z_{k/k-1})^{\rm T}(\Sigma_{w}^{-1/2}{\rm C_{1}}Z_{k/k-1})]^{-1}(\Sigma_{w}^{-1/2}{\rm C_{1}}Z_{k/k-1})^{\rm T}\Sigma_{w}^{-1/2}(y_{k} - \overline{y}_{k/k-1})\}.
\end{aligned}
\end{equation}
\indent Let us define the vector $S_{inov}= \Sigma_{w}^{-1/2}(y_{k} - \overline{y}_{k/k-1})$ and the matrix $S_{cz} = \Sigma_{w}^{-1/2}{\rm C_{1}}Z_{k/k-1}$, then Eq. \eqref{Est_majA2} becomes:
\begin{equation}\label{Est_majA3}
\hat{x}_{k/k} = \overline{x}_{k/k-1} + Z_{k/k-1}S_{cz}^{\rm T}\{S_{inov}-S_{cz}[{\rm Id} + S_{cz}^{\rm T}S_{cz}]^{-1}S_{cz}^{\rm T}S_{inov}\}.
\end{equation}
\indent By using the EVD of the matrix $({\rm Id} + S_{cz}^{\rm T}S_{cz})$ (which gives the expression of the ETM $T_{k}$ in Eq. \eqref{Mat_T_QG}), we obtain a new decomposition for the matrix inversion in the square brackets of last Eq. \eqref{Est_majA3}: therefore, it does not require the inversion of the $p\times p$ matrix in the brackets of the original Eq. \eqref{K_Gain}, but only a $m\times m$ matrix EVD which can be computationally cheaper if $m \ll p$. We finally obtain for the update estimate:
\begin{equation}\label{Est_majA4}
\hat{x}_{k/k} = \overline{x}_{k/k-1} + Z_{k/k-1}S_{cz}^{\rm T}\{S_{inov}-S_{cz}Q_{k}\Gamma_{k}^{-1}Q_{k}^{\rm T}S_{cz}^{\rm T}S_{inov}\}.
\end{equation}

\section*{Appendix B: Total number of operations for the ETKF on a zonal basis}

The mathematical formalism presented in this paper is valid for both a modal or a zonal basis. But using a zonal basis (where the phase is estimated on the locations of each valid actuator of the DM) enables to compute some very sparse matrices which is very suitable for calculations on large-scale AO systems. Let us define $\rm n_{act}$ the number of valid actuators: on a zonal basis with an AR1 turbulence model in the ETKF-based control law, as $\rm A_{tur} = \rm a_{tur}\times Id$, the extracted matrix $\rm A_{1} = \left(\begin{matrix} {\rm A_{\rm tur}} & 0 \\ {\rm Id} & 0 \\ \end{matrix}\right)$ is composed by only two $\rm n_{act}\times n_{act}$ diagonal blocks. Thus, $\rm A_{1}$ is a very sparse matrix and its size is $\rm n\times n$, where $\rm n = 2\times\rm n_{act}$. Let us define $\rm p_{sap}$ the number of valid subapertures of the S-H WFS (each subaperture gives two slopes of the residual phase in two directions): the dimension of the measurement vector $y_{k}$ is $\rm p\times 1$, where $\rm p = 2\times\rm p_{sap}$. Let us define the matrix $\rm D_{1}$, modeling the S-H WFS on the zonal basis: this matrix $\rm D_{1}$ is the equivalent of the linear operator D in Eq. \eqref{Eq_Obs}. For a SH-WFS with a Fried geometry, $\rm D_{1}$ enables to calculate 2 slopes (at the center of each subaperture) from the 4 estimations of the turbulent phase on the actuators at the 4 corners of each subaperture: its size is $\rm p\times n_{act}$ and each row of this matrix is then composed by only 4 non-zero values. Thus, $\rm D_{1}$ is a very sparse matrix. Moreover the extracted observation matrix $\rm C_{1} = \left[\begin{matrix} 0 & {\rm D_{1}} \end{matrix} \right]$ is also very sparse and its size is $\rm p\times n$. The influence matrix N characterises the DM on the zonal basis by Eq. \eqref{Eq_Cor}: its size is $\rm n_{act}\times n_{act}$ and it is a very sparse matrix. For the calculation of $\overline{y}_{k/k-1}$, we have to compute ${\rm D_{1}N}\times u_{k-2}$ where the matrix ${\rm D_{1}N}$ is the result of the mulplication of 2 sparse matrices. In our classical AO configuration, for a 16 m, a 32 m and a 40 m telescope, the number of non-zero values per row of this matrix ${\rm D_{1}N}$ is always less than 36.

\subsection*{B.1 The Update step}

By using the last expression (Eq. \eqref{Est_majA4}) and the associativity property of matrix multiplication, we can compute many matrix-vector multiplications in order to minimize as far as possible the theoretical numerical cost.\\
\indent  Actually the significant computational expense is the EVD and the 3 matrix-matrix multiplications: $S_{cz}^{\rm T}\times S_{cz}$ , $Q_{k}\times(\sqrt{m-1}\Gamma_{k}^{-1/2}Q_{k}^{\rm T})$ and $Z_{k/k-1}\times(Q_{k}\sqrt{m-1}\Gamma_{k}^{-1/2}Q_{k}^{\rm T})$.
\begin{table}[H]
\centering\caption{Numbers of multiplications during the update step.}
\begin{tabular}{|c|c|c|}
\hline
Expressions & Multiplications & Result Size \\
\hline
$\overline{y}_{k/k-1} = {\rm C_{1}}\times\overline{x}_{k/k-1} - {\rm D_{1}N}\times u_{k-2}$ & $ p\times 4 + p\times 36$ & (p,1) \\
$S_{inov} = \Sigma_{w}^{-1/2}\times(y_{k} - \overline{y}_{k/k-1})$ & $p$ & (p,1) \\
$S_{cz} = \Sigma_{w}^{-1/2}{\rm C_{1}}\times Z_{k/k-1}$ & $p\times 4\times m$ & (p,m) \\
$S_{cz}^{\rm T}\times S_{cz}$ & $m\times p\times m$ & (m,m) \\
EVD of $({\rm Id} + S_{cz}^{\rm T}S_{cz})$ & $m^{3}$ & (m,m) \\
$\Gamma_{k}^{-1}\qquad$ and $\qquad\Gamma_{k}^{-1/2}$& $m + \gamma\times m$ & (m,m) \\
$S_{cz}^{\rm T}\times S_{inov}$ & $m\times p$ & (m,1) \\
$S_{cz}\times(Q_{k}\times(\Gamma_{k}^{-1}\times(Q_{k}^{\rm T}\times(S_{cz}^{\rm T}S_{inov}))))$ & $m^{2} + m + m^{2} + p\times m$ & (p,1) \\
$S_{cz}^{\rm T}\times(S_{inov}-S_{cz}Q_{k}\Gamma_{k}^{-1}Q_{k}^{\rm T}S_{cz}^{\rm T}S_{inov})$ & $m\times p$ & (m,1)\\
$Z_{k/k-1}\times S_{cz}^{\rm T}(S_{inov}-S_{cz}Q_{k}\Gamma_{k}^{-1}Q_{k}^{\rm T}S_{cz}^{\rm T}S_{inov})$ & $n\times m$ &  (n,1) \\
$Z_{k/k-1}\times(Q_{k}\times((\sqrt{m-1}\times\Gamma_{k}^{-1/2})\times Q_{k}^{\rm T}))$ & $m + m^{2} + m^{3} + n\times m\times m$ & (n,m) \\
\hline
\end{tabular}
\end{table}
Total number of multiplications: $(m^{2}+m)\times n + (m^{2}+7m+41)\times p + 2m^{3}+3m^{2}+(3+\gamma)m$ 
The number of additions is the same order of magnitude as the number of multiplications.

\subsection*{B.2 The Prediction step}

By using a zonal basis with an AR1 turbulence model, $\rm A_{1}$ is composed by two $n_{act}\times n_{act}$ diagonal blocks, one of them is the identity matrix. Therefore, multiplying $\rm A_{1}$ with a vector is reduced to only one multiplication with the block $\rm A_{\rm tur}$ (the other one consists on a copy).
\begin{table}[H]
\centering\caption{Numbers of operations during the prediction step.}
\begin{tabular}{|c|c|c|c|}
\hline
Expressions & Multiplications & Additions & Result Size \\
\hline
$X_{k+1/k} = {\rm A_{1}}\times X_{k/k} + {V}_{k+1}$ & $\frac{n}{2}\times m$ & $\frac{n}{2}\times m$ & (n,m) \\
$\overline{x}_{k+1/k} = \frac{1}{m}\sum\limits_{i = 1}^{m}{x}_{k+1/k}^{i}$ & $n$ & $n\times(m-1)$ & (n,1) \\
$Z_{k+1/k} = \frac{[x_{k+1/k}^{1}-\overline{x}_{k+1/k};...;x_{k+1/k}^{m}-\overline{x}_{k+1/k}]}{\sqrt{m-1}}$ & $n\times m$ & $n\times m$ & (n,m) \\
\hline
\end{tabular}
\end{table}
Total number: $(\frac{3}{2}m + 1)\times n$ multiplications + $(\frac{5}{2}m - 1)\times n$ additions 

\subsection*{B.3 The Projection onto the DM}

The voltage $u_{k}$ is calculated with the MVM: $u_{k} = {\rm (N^{T}N)^{-1}N^{T}}\times\hat{\varphi}_{k+1/k}^{\rm tur}$.\\
The matrix $\rm P = (N^{T}N)^{-1}N^{T}$ is a sparse matrix on a zonal basis. In our SCAO system, for a 16 m (respectively a 40 m) telescope, the sparsity of this matrix is 51 \% (respectively 88 \%). Actually, on a zonal basis, the matrix P can be much more sparse when, for each actuator, we take into account only the neighboring actuators close to less than 2 pitches.


\section*{Acknowledgments}

This work has been supported with financial grants from the cross-disciplinary mission MASTODONS of the CNRS and from the Programme Hubert Curien-AURORA (Campus France) for mobilities between France and Norway. 


\end{document}